\documentclass[aps,prab,groupedaddress,floats,reprint]{revtex4-2}

\usepackage[english] {babel}
\usepackage{amsfonts,amssymb,graphicx,soul,natbib,mathrsfs,dsfont}

\usepackage{epstopdf}
\usepackage{xcolor}
\usepackage{soul}
\usepackage{physics}
\usepackage{graphics}
\usepackage{comment}

\renewcommand{\vec}[1]{\textnormal{\boldmath$#1$}}

\usepackage{tikz}
\usepackage{amscd}
\usepackage[inline]{enumitem}
\usepackage[
    left=1.9cm,
    right=1.9cm,
    top=1.5cm,
    bottom=2cm,
    bindingoffset=0cm,
]{geometry}
\usepackage{amsmath, nicefrac}

\usepackage{cancel}
\usepackage{physics}

\usepackage[normalem]{ulem}

\usepackage{dcolumn}
\usepackage{bm}

\newcommand{\rd}[1]{\textcolor{black}{#1}}
\newcommand{\bl}[1]{\textcolor{black}{#1}}

\usepackage[colorlinks=true]{hyperref}

\begin{document}

\title{Analytical theory of the skewed wake effect}

\author{A. N. Chuprina}%
\email{anna.chuprina@metalab.ifmo.ru}%
\affiliation{School of Physics and Engineering,
ITMO University, St. Petersburg, Russia 197101}%

\author{S. S. Baturin}%
\email{s.s.baturin@gmail.com}%
\affiliation{School of Physics and Engineering,
ITMO University, St. Petersburg, Russia 197101}%

\date{\today}
\begin{abstract}
Recently, the experimental discovery of a new type of wakefield effect, the ``skewed wake effect", has been reported. We provide an explanation of the nature of the skewed wake effect based on a simple three-particle model that we have developed. Taking a step forward, we analyze this effect for the case of a highly elliptical beam, provide a simple estimate of the skew angle, and analyze the wake amplitude effects associated with this effect.
\end{abstract}

\maketitle

\section{INTRODUCTION}

The Panofsky-Wentzel theorem \cite{PW} states that any asymmetry in the transverse distribution of the longitudinal electric field will inevitably lead to the appearance of the transverse component of the Lorentz force, which in turn results in emittance growth and may even cause beam break-up instability \cite{Chao,Li:2014,BBU2,BBU3,Obzor}.

Such interactions between particles within a beam or between different beams (e.g., driver and witness) through radiation are typically described within a convenient framework known as wakefield interactions \cite{BaneCol,HfKf,Chao,Zotter}. To date, numerous types of wakefields have been identified and analyzed, such as geometric wakefields, resistive wall wakefields \cite{Reswk1,Dimp,Reswk3}, dielectric wakefields \cite{NG,Gai,Qd1,mySTAB}, metamaterial wakefields \cite{Andonian:2014,Hoang:2018} \bl{and plasma wake fields \cite{Rosen1991,Mora1997,Lotov2004,Barov,Lu,Stupakov:2016,Stupakov:2018,Mori}}.

For conventional machines like CLIC \cite{CLIC}, ILC \cite{ILC}, LCLS-II \cite{LCLS}, and especially for wakefield accelerators like Bella \cite{Bella}, FlashFORWARD \cite{FLASHfwd}, and A-STAR \cite{Zhol,Zhol2,Zhol3}, the consideration of wakefield effects is crucial. In collider beams, beam-beam interactions can negatively affect luminosity \cite{Chen}. \rd{In circular machines such as synchrotrons when the beam intensity and/or brightness is pushed the wake field effects become crucial as they triggers different types of collective instabilities.\cite{SYL}}

Recently, it has been shown \cite{planar,Brendan:2020} that wakefield interactions depend significantly on the geometry of the problem. For example, a highly elliptical beam generates significantly weaker transverse wakefields \cite{FLb,planar}. The concept of driver asymmetry has found applications in hollow-channel plasma \cite{Joshi} and has emerged in proposals for flat-beam plasma accelerators \cite{FlatPlasma}. For collider applications, K. Yokoya and P. Chen argued that highly elliptical beams could be used to suppress bremsstrahlung at the interaction point \cite{Chen2}.

The production of highly asymmetric beams and methods for refocusing at the collider interaction point have been extensively studied. To date, an emittance ratio of  $\varepsilon_x/\varepsilon_y\sim400$ (and ellipticity $\sigma_x/\sigma_y >20$) \cite{Piot2,Piot2c} has been successfully demonstrated. Thus, flat-beam technology is already reasonably well-developed and justified.

Regarding geometry, one might intuitively conclude that not only the shape of the guiding channel (e.g., round or slab) matters, but also the shape of the driver and its relative position within the guide channel. This is especially true if one of the transverse dimensions of the beam is comparable to the minimum characteristic size of the vacuum channel. This intuition has recently been experimentally confirmed. In Ref.\cite{SKWK}, the authors report the discovery of the ``skew\rd{ed} wake effect", which manifests as a rotation of the dipole and quadrupole wake toward the beam when the beam is tilted in the transverse plane relative to the boundary of the guiding slab channel. This discovery and the detailed study of this effect represent a critical step in understanding the prospects and limitations of the flat-beam concept.

In the present paper, we study the quadrupole skew wake effect from a theoretical perspective. Our analysis is based on the transverse Green's function estimation introduced in Ref.\cite{MyPRL,mySTAB,planar}. We start from the model presented in Ref.\cite{planar} and develop a three-particle model, which reveals the essence of the skew wake field mechanism. We also discuss important peculiarities, such as the dependence of the magnitude of the skew wake effect on the asymmetry of the driver. We then develop a model of a highly elliptical Gaussian beam and derive an analytical estimate of the skew angle of the skew quadrupole wake. Interestingly, we find that the skew angle is 1.5 times the transverse tilt angle of the beam \rd{with a minus sign, i.e. the wake rotates towards the beam}.

\section{Physics Mechanism of the Skew Wake Effect} \label{section with general formulas}

The theoretical formula for the upper limits of the transverse component of the Lorentz force in the slab channel, \( F_\perp \), was derived in Ref.~\cite{mySTAB,planar} and reads
\begin{align}
\label{eq:Ftr}
        F_{\perp}(\psi,\psi_0) = -\dfrac{4qQ\pi^3\theta(\zeta)\zeta}{32a^3}\frac{\tanh\left[\dfrac{\pi(\psi^*-\psi_0)}{4a}\right]}{\left\{\cosh\left[\dfrac{\pi(\psi^*-\psi_0)}{4a}\right]\right\}^2}, 
\end{align}
where \( Q \) is the charge of the particle generating the wakefield, \( q \) is the charge of the test particle, \( a \) is the size of the structure aperture, \( \zeta = ct - z \) is the longitudinal distance between the particle driving the wakefield and the test particle, \( \theta(\zeta) \) is the Heaviside function (which ensures that the field is nonvanishing only behind the driving particle), \(\psi = x+i y \) are the coordinates of the test particle, and \( \psi_0 = x_{0}+ i y_{0} \) are the coordinates of the particle generating the wakefield in the cross-section of the structure. Above and further in the text, asterisks stand for complex conjugation.

It is convenient to introduce a normalized wake potential per unit length as 
\begin{align}
    w_\perp = F_{\perp}  \frac{8a^3}{qQ \pi^3 \theta(\zeta)\zeta}.
\end{align}
    
With this normalization, from \eqref{eq:Ftr}, we obtain
\begin{equation} \label{potent}
\begin{aligned}
w_{\perp}(x,y,x_0,y_0) = -\frac{\tanh\left[\dfrac{\pi(\psi^*-\psi_0)}{4a}\right]}{\left\{\cosh\left[\dfrac{\pi(\psi^*-\psi_0)}{4a}\right]\right\}^2}.
\end{aligned}
\end{equation}

\begin{figure}[t]
    \centering
   \includegraphics[width=0.49\textwidth]{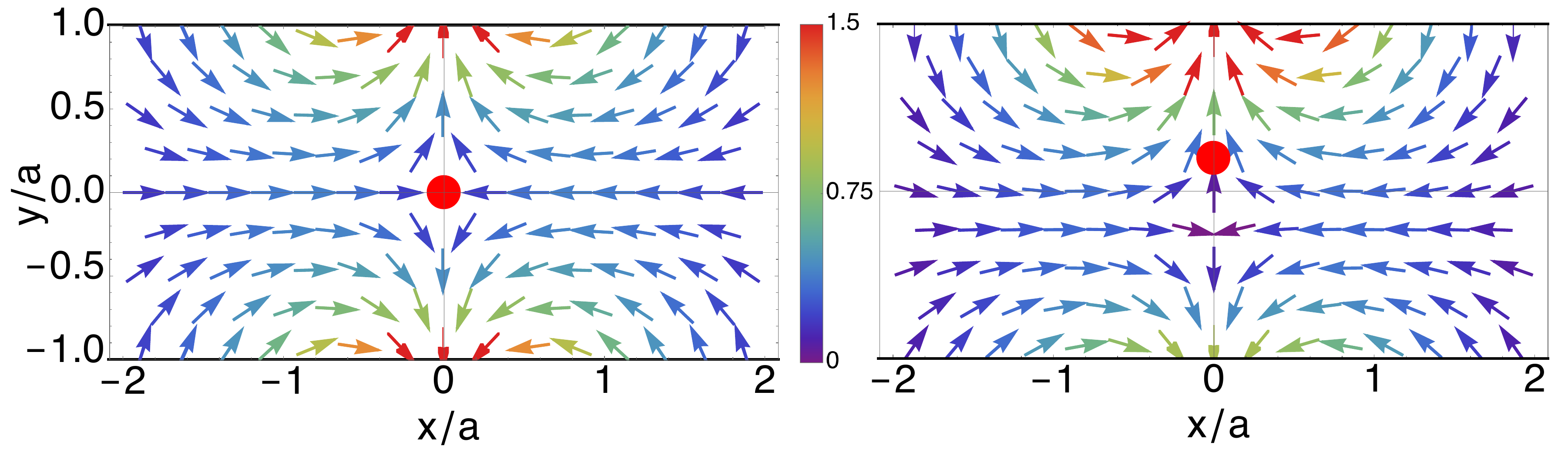}
    \caption{Field lines of the transverse wake potential given by Eq. (\ref{potent}). The point particle is schematically represented by a red circle. In the left panel, the point particle is located at the center of the wakefield structure; in the right panel, the point particle is shifted from the center to the point \( y_0 = 0.2a \).}
    \label{point particle}
\end{figure}

Let us examine the field lines for the transverse wake potential (\ref{potent}), which is created by a point particle. Consider a particle moving along the axis of a planar structure, as shown in Fig.~\ref{point particle} (left panel), and a particle that is displaced by \( y_0 = 0.2a \) from the center, as shown in Fig.~\ref{point particle} (right panel).

Note that the transverse wake potential has a quadrupole structure in the region \( |x/a|<1 \) and \( |y/a|<1 \), and a vortex-like structure further away from this region. When the point particle is moved upward, the amplitude of the transverse wake field increases near the upper boundary and decreases at the lower boundary. This simple observation leads to an intuitive statement (which can be seen directly from Eq.~\eqref{potent}): the closer the particle is to the boundary, the higher the amplitude of the transverse wake potential. 

The second observation from the figure is that the structure of the wake is always \rd{quadrupolar}, with the vertical axis passing straight through the source particle. The third observation is that the origin of this quadrupole shifts in the direction perpendicular to the boundary. This shift is always opposite to the shift of the source. In Fig.~\ref{point particle} \bl{(right panel)}, the particle is shifted upward, and the center of the quadrupole wake is shifted downward. This behavior can be well explained from the perspective of multipole decomposition. Near the origin, the wake consists of two main components: a quadrupole (which is dominant) and a dipole (which appears only when the particle is displaced toward the wall) \cite{BaneNima}:
\begin{align}
w_\perp \approx \underbrace{i\frac{\pi}{4a}y_0}_{\text{dipole}} + \underbrace{-\frac{\pi}{4a}x + i\frac{\pi}{4a}y}_{\text{quadrupole}}.
\end{align}
Above, \( y_0 \) is the displacement of the source particle in the \( y \)-direction, and \( x \) and \( y \) are the coordinates of the test particle.

\begin{figure}[t!]
    \centering
   \includegraphics[width=0.49\textwidth]{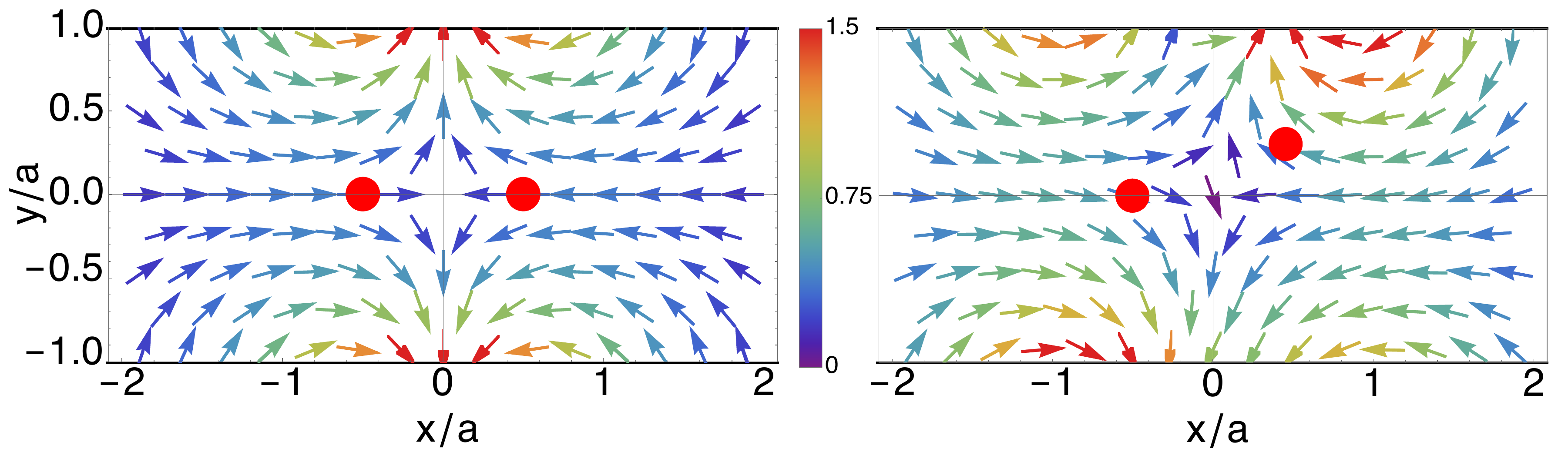}
    \caption{Field lines of the transverse wake potential given by Eq. (\ref{potent2}). The point particles are schematically represented by red circles. In the left panel, the point particles are located at the center of the wakefield structure; in the right panel, one of the point particles is rotated relative to the other by an angle \( \pi /10 \). The distance between the particles is \( L = 0.5a \).}
    \label{two particles}
\end{figure}

Next, we consider the case where the wakefield is generated by two point particles. The transverse wake potential is simply the sum of the individual wake potentials of each particle and reads
\begin{equation} \label{potent2}
    w_{\perp, 2}(x,y) = w_{\perp}(x,y,-L,0) + w_{\perp}(x,y,L,0).
\end{equation}
Here, \bl{\( 2L < a \)} is the transverse \( x \)-distance between the particles.

In Fig.~\ref{two particles}, we plot the field lines of the transverse wake potential (\ref{potent2}) for two cases: when the two point particles are traveling along the center of the structure (Fig.~\ref{two particles}, left panel) and when one point particle is shifted toward the boundary (Fig.~\ref{two particles}, right panel). For the symmetric case, the structure of the transverse wake potential in the region \( |y|<a \), \( |x|<a \) resembles the same quadrupole as in the single-particle case. However, in the case where the second particle is shifted toward the boundary, the situation is different.

We note that the wakefield generated by the second particle (the one closer to the boundary) is higher than the wakefield generated by the first one. Thus, the sum exhibits an asymmetry; i.e., the vector field in the upper plane ($1 > y/a > 0$) is ``pulled" toward the \rd{``hot spot"} at the boundary created by the second particle. \rd{At the same time, in the lower plane ($0 > y/a > -1$)}, the field of the second particle is weakened, and the resulting vector field is attracted to the \rd{``hot spot"} at the boundary created by the first particle.

The total asymmetry of the resulting wake is observed as a deformation of the quadrupole structure, which can be decomposed into the rotation of the principal axis of the quadrupole, amplitude scaling, and a stretch. The latter is connected with the dipole component of the wake potential, which is generated due to the displacement of the center of mass toward the upper boundary. In the case where the center of mass of the two particles is located at the origin, the quadrupole wake experiences only rotation and amplitude scaling.

We note that the rotation of the principal axis of the quadrupole always occurs toward the rotation of the beam. From the two-particle example, it is clear that this happens due to the dipole component of the wake generated by the first and second particles, which have the same magnitude but opposite signs.

We point out that the skew effect is a result of the interplay between the dipole wakes generated by different parts of the beam.

\section{The Three-Particle Model} \label{section:three-particle-model}

The simplest and most informative model of a transversely asymmetric beam is the three-particle model, where the central particle represents the core of the beam, and the two satellite particles reflect significant asymmetry in the beam distribution along one direction.

We start with the transverse wake potential of the three particles, \( g_{\perp}(x,y) \), in the form
\begin{multline}\label{potential-3-part-no-rotate}
     g_{\perp}(x,y) = w_{\perp}(x,y,-d,0) + w_{\perp}(x,y,0,0) \\+ w_{\perp}(x,y,d,0),
\end{multline}
where \( d \) represents the distance between the particles, and \( w_\perp \) is given by Eq.~\eqref{potent}.

Next, we assume that the central particle remains at the origin, while the two satellite particles rotate around the origin at the same angle \( \alpha \). The expression for the transverse wake potential \( g_{\perp}(x,y) \) then becomes
\begin{multline} 
\label{potent-3-particle-rotate}
    g_{\perp}(x,y) = w_{\perp}(x,y,-d \cos \alpha, -d \sin \alpha)\\ + w_{\perp}(x,y,0,0) + w_{\perp}(x,y,d \cos \alpha, d \sin \alpha).
\end{multline}
We then linearize the transverse wake potential \( g_{\perp}(x,y) \) near the origin and examine the field created by the three rotated particles.

We decompose \( g_{\perp}(x,y) \) into a Taylor series with respect to \( x \) and \( y \) around the origin, keeping only the linear terms:
\begin{multline} \label{G-S}
    g_0^S(x,y) = \pi\frac{-x+i y}{4a} 
    \frac{1+2\left[2-\cosh \left(\frac{1}{2a}\pi  e^{i \alpha } d \right) \right]}{\left[\cosh \left(\frac{1}{4a} \pi  e^{i \alpha } d \right)\right]^4}.
\end{multline}
Here, the index \( S \) stands for the general skewed case where \( \alpha \neq 0 \). \bl{Note that the real and imaginary parts ($g_x\equiv\Re[g]$ and $g_y\equiv \Im[g]$) of the function $g$ are the $x$ and $y$ components of the transverse wake potential, respectively.}

We note that \( g_0^S(x,y) = g_x^S(x,y) + i g_y^S(x,y) \) can be expressed in terms of the regular quadrupole wake potential \( g_0(x,y) = g_x(x,y) + i g_y(x,y) \), created by the three particles aligned along the \( x \)-axis (a special case of \bl{Eq.~\eqref{G-S}} when \( \alpha = 0 \)):
\begin{align} \label{G-0}
    g_0(x,y) = \pi\frac{-x+i y}{4a} \frac{1+2\left[2-\cosh \left(\frac{1}{2a}\pi d \right) \right]}{\left[\cosh \left(\frac{1}{4a} \pi d \right)\right]^4}.
\end{align}
The connection between the skewed wake potential \( g_0^S(x,y) \) and the normal wake potential \( g_0(x,y) \) can be conveniently expressed in matrix form:
\begin{equation} \label{matrixes}
\begin{aligned}
 \begin{pmatrix}
  g_x^S\\
  g_y^S
\end{pmatrix} = 
A\begin{pmatrix}
  g_x\\
  g_y
\end{pmatrix},
\end{aligned}
\end{equation}
where \( A \) is the modified normalized Jacobian matrix given by
\begin{align}
A = \frac{J}{N} = \frac{1}{N}\begin{pmatrix}
 -\partial_x \mathrm{Re}[g_\perp] & \partial_y \mathrm{Re}[g_\perp]\\
  -\partial_x \mathrm{Im}[g_\perp]& \partial_y \mathrm{Im}[g_\perp]
\end{pmatrix}.
\end{align}
Above partial derivatives are \rd{taken} at the point $x=y=0$.
The normalization coefficient is
\begin{align}
N = \frac{\pi}{4a} \frac{1+2\left[2-\cosh \left(\frac{1}{2a}\pi d \right) \right]}{\left[\cosh \left(\frac{1}{4a} \pi d \right)\right]^4},
\end{align}
and the signs in the first column of the matrix \( J \) are chosen such that when \( \alpha = 0 \), the matrix \( J \) becomes an identity matrix. Using the analytic properties of the \( g_\perp \) function \cite{planar,BaneNima,shabat,silverman}, we obtain
\begin{align}
&\partial_x \mathrm{Re}[g_\perp] = -\partial_y  \mathrm{Im}[g_\perp], \nonumber\\
&\partial_y \mathrm{Re}[g_\perp] = \partial_x  \mathrm{Im}[g_\perp].
\end{align}
From these expressions, it follows that \( J \) is non-degenerate, i.e.,
\begin{align}
\mathrm{det} J = \left(\partial_y  \mathrm{Im}[g_\perp] \right)^2 + \left(\partial_x  \mathrm{Im}[g_\perp] \right)^2 \neq 0.  
\end{align}

\begin{figure}[t]
    \centering
   \includegraphics[width=0.49\textwidth]{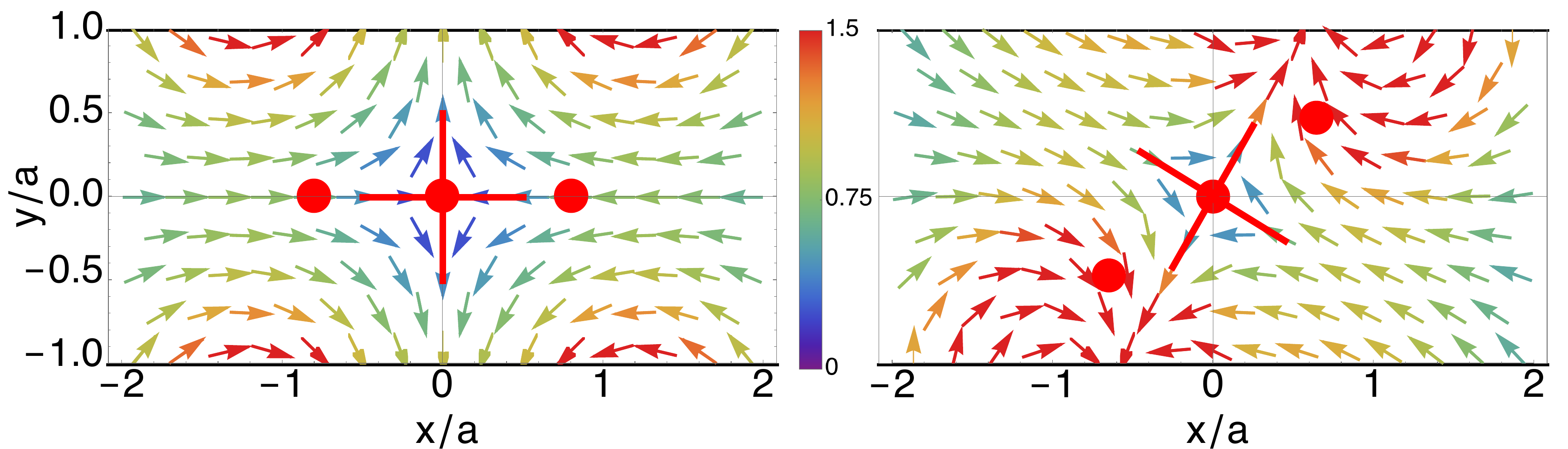}
    \caption{Field lines of the transverse wake potential given by Eq.~\eqref{potential-3-part-no-rotate} (left panel) and Eq.~\eqref{potent-3-particle-rotate} (right panel). The point particles are schematically represented by red circles. In the right panel, the rotation angle for the particles is taken to be \( \pi /5 \). The distance between the particles is \( d = 0.8a \). The red lines are calculated as \( y_1 = x \tan \phi \) and \( y_2 = -x\cot \phi \), where \( \phi \) is given by Eq.~\eqref{value-of-the-angle-separatrices}.}
    \label{fig:3}
\end{figure}

We introduce another normalization and express the matrix \( A \) as
\begin{align}
\label{eq:transf}
A = \frac{\sqrt{\mathrm{det} J}}{N}O,
\end{align}
where \( O \) is a rotation matrix with the following form:
\begin{align}
\label{eq:rot}
O = \begin{pmatrix}
 \frac{\partial_y \mathrm{Im}[g_\perp]}{\sqrt{\mathrm{det}J}} & \frac{\partial_x \mathrm{Im}[g_\perp]}{\sqrt{\mathrm{det}J}}\\
  -\frac{\partial_x \mathrm{Im}[g_\perp]}{\sqrt{\mathrm{det}J}}& \frac{\partial_y \mathrm{Im}[g_\perp]}{\sqrt{\mathrm{det}J}}
\end{pmatrix}.
\end{align}
From Eq.~\eqref{eq:transf}, it immediately follows that the linearized transverse skewed wake potential is simply the regular quadrupole transverse wake potential, scaled by a factor of \( \sqrt{\mathrm{det} J}/N \) and rotated by an angle
\begin{align} \label{value-of-the-angle-separatrices}
    \phi = \frac{1}{2} \arctan \left( \dfrac{\partial_x \mathrm{Im}[g_\perp]}{\partial_y \mathrm{Im}[g_\perp]} \right).
\end{align}
Details on the coefficient \( 1/2 \) are provided in Appendix~\ref{app:separ}.

By inspecting Eq.~\eqref{G-S}, we observe that it is convenient to express the transformation \eqref{eq:transf} in terms of the complex variable \( N^S \):
\begin{align}
N^S = \frac{\pi}{4a}\frac{1+2\left[2-\cosh \left(\frac{1}{2a}\pi  e^{i \alpha } d \right) \right]}{\left[\cosh \left(\frac{1}{4a} \pi  e^{i \alpha } d \right)\right]^4}.
\end{align}
In this case, the rotation angle for the skewed wake becomes
\begin{align}
\label{eq:skang}
\phi = \frac{\mathrm{arg} [N^S]}{2}
\end{align}
and the scaling factor is
\begin{align}
\label{eq:skamp}
\lambda = \frac{\sqrt{\mathrm{det}J}}{N} = \frac{|N^S|}{N}.
\end{align}
To illustrate Eq.~\eqref{value-of-the-angle-separatrices} and Eq.~\eqref{eq:skang}, we plot the transverse distribution of the wake potential for the tilted beam (Fig.~\ref{fig:3}, right panel). We observe that the cross corresponding to the separatrices of the quadrupole wake is rotated and reproduces the lines of the fastest descent of the field quite well.

\begin{figure}
    \centering 
    \includegraphics[width=1\linewidth]{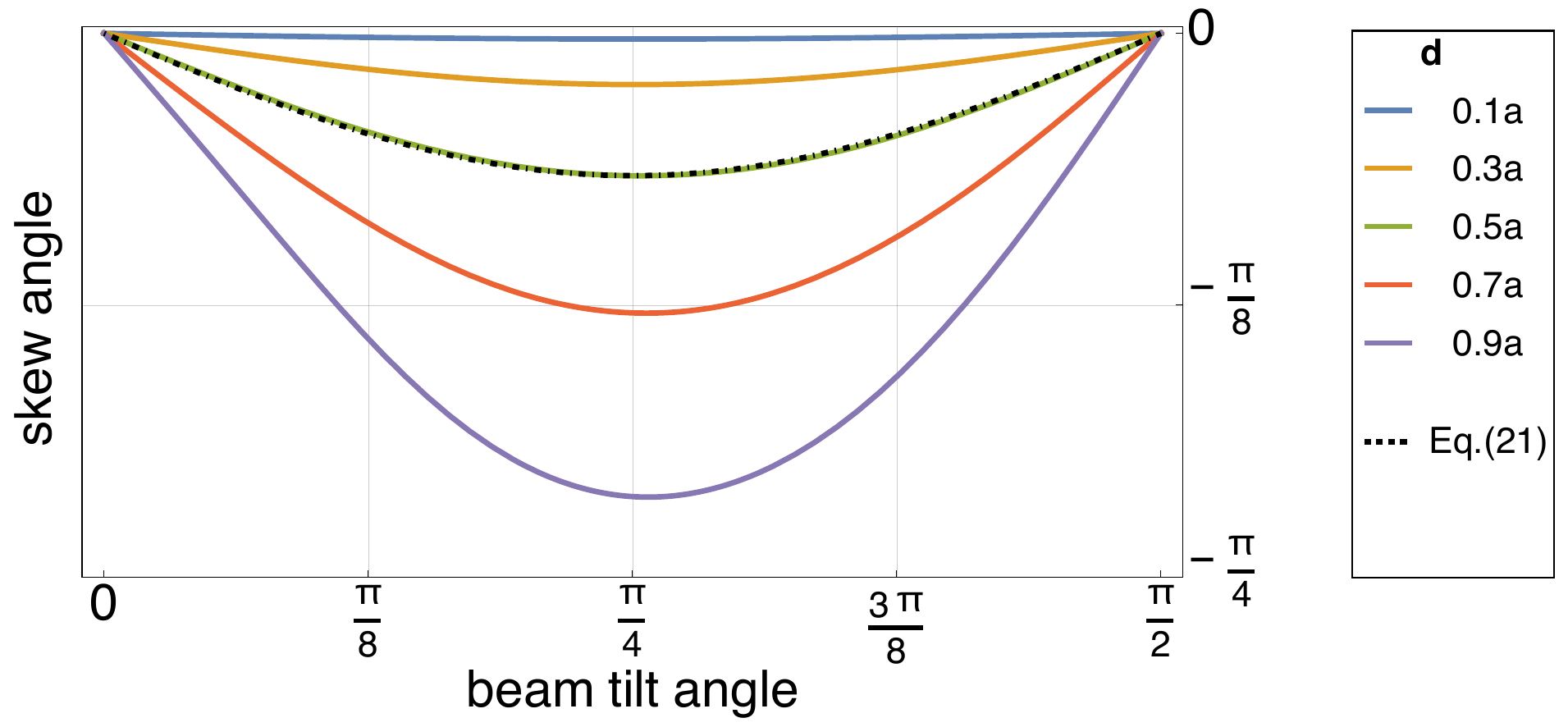}
    \caption{Quadrupole skew wake angle \( \phi \) given by Eq.~\eqref{eq:skang} as a function of the beam tilt angle \( \alpha \) for different values of the width parameter \( d \). The dashed line represents the approximate formula for the tilt angle given by Eq.~\eqref{eq:approxsk} \rd{for $d =0.5$}.}
    \label{fig:4}
\end{figure}

Next, we analyze the dependence of the skew angle (Eq.~\eqref{eq:skang}) and the scaling coefficient (Eq.~\eqref{eq:skamp}) on the half-width \( d \) and the tilt angle \( \alpha \) of the three-particle beam. In Fig.~\ref{fig:4}, we plot the dependence of the skew angle on the tilt angle for five different values of \( d \). We observe that the skew effect (i.e., the rotation of the quadrupole wake) is insignificant for a symmetric beam with \( d \sim 0.1a \). However, when the full width approaches the size of the aperture (\( 2d \sim a \)), the effect becomes pronounced. Up to a tilt angle of \( \pi/4 \), the \rd{absolute value of the} skew angle grows to a maximum value that depends on the beam width and then drops back to zero, as in the limiting case of \( \alpha = \pi/2 \), the symmetry is restored, and all three particles are positioned along the \( y \)-axis.

To gain insight into the amplitude of the skew angle, we expand Eq.~\eqref{eq:skang} in a series with respect to \( d \), keeping only the first significant term:
\begin{align}
\label{eq:approxsk}
\phi \approx -\frac{\pi^2 d^2}{12a^2}\sin 2\alpha.
\end{align}

\begin{figure} [t]
    \centering
    \includegraphics[width=1\linewidth]{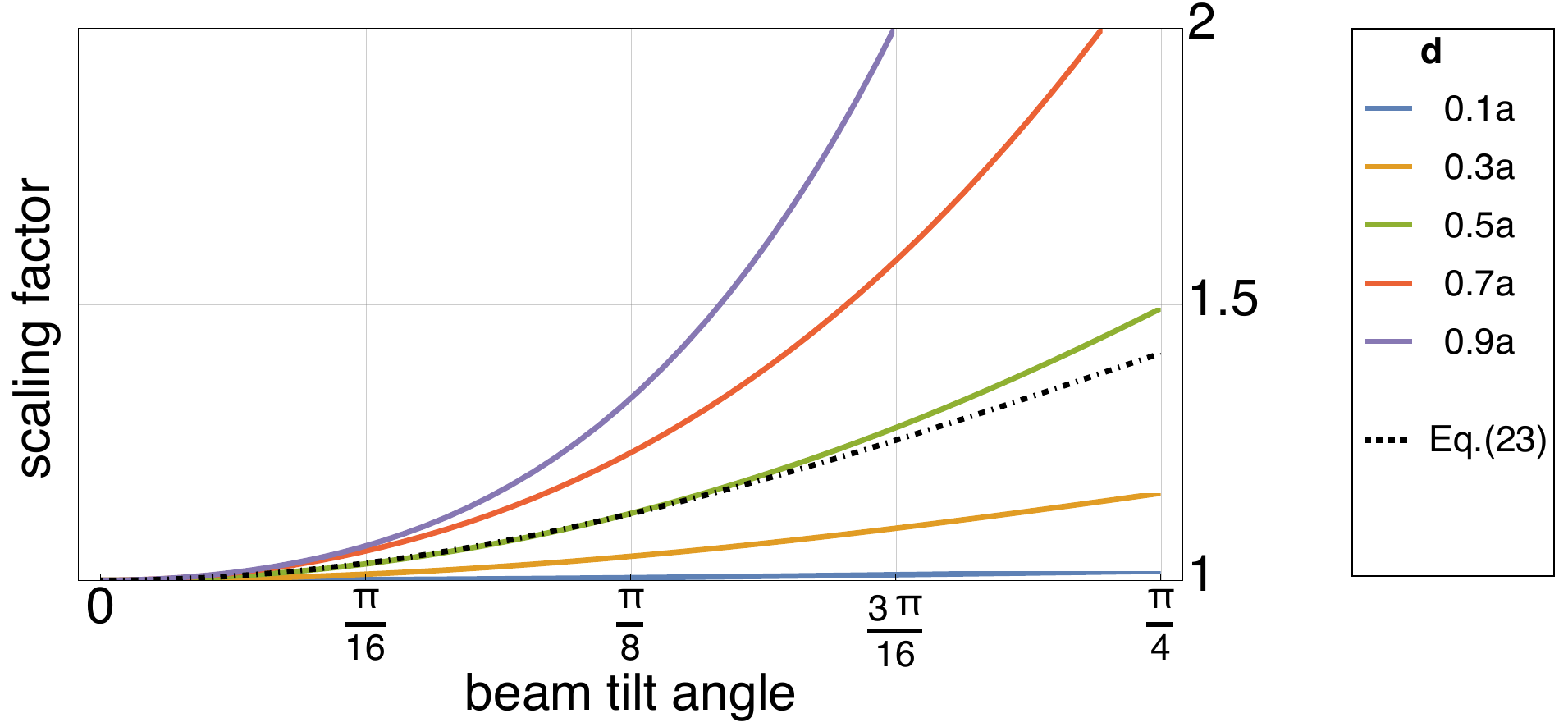}
    \caption{Wake scaling coefficient \( \lambda \) given by Eq.~\eqref{eq:skamp} as a function of the beam tilt angle \( \alpha \) for different values of the width parameter \( d \). The parameter \( d \) is measured in units of the half-aperture \( a \). \rd{The dashed line represents the approximate formula for the
scaling coefficient \( \lambda \) given by Eq.~\eqref{eq:approxap} for $d =0.5$}.}
    \label{fig:5}
\end{figure}

We note the negative sign, which indicates that the skew wake rotates in the opposite direction to the rotation of the beam. The above equation works well for \( d \leq 0.5 \), as can be seen in Fig.~\ref{fig:4}, where Eq.~\eqref{eq:approxsk} is plotted with a dashed line for the case of \( d = 0.5 \). From Eq.~\eqref{eq:approxsk}, we immediately conclude that the \rd{peak} value of the skew angle is
\begin{align}
\phi_{\mathrm{peak}} \approx -\frac{\pi^2 d^2}{12a^2}.
\end{align}
An approximate expression for the scaling parameter is derived from Eq.~\eqref{eq:skamp} in the same manner and reads
\begin{align}
\label{eq:approxap}
\lambda \approx 1 + \frac{\pi^2d^2}{3a^2}\left(\sin \alpha \right)^2.
\end{align}
This equation works well for the case of small \( d \leq 0.5 \) and \( \alpha \leq \pi/8 \), as illustrated in Fig.~\ref{fig:5}, where Eq.~\eqref{eq:approxap} is plotted with a dashed line for the case of \( d = 0.5 \).

In the limit of small tilt angles, we obtain
\begin{align}
\phi \approx -\frac{\pi^2 d^2}{\bl{6}a^2}\alpha,
\end{align}
and
\begin{align}
\lambda \approx 1 + \frac{\pi^2d^2}{3a^2}\alpha^2.
\end{align}
The latter equation indicates that for \( \alpha < 0.25 \), the rotation effect is dominant. We point out that the scaling factor \( \lambda \) grows rapidly, and for an asymmetric beam with \( d = 0.7 \), the increase in the transverse wake amplitude can reach \( \sim 32\% \) for a beam tilt angle of \( \pi/8 \) (see Fig.~\ref{fig:5}). However, as we show in the next section, this effect is significant only for beams with a full width on the order of \( \sim a \).

It is worth reiterating that the skew wake effect is twofold (as one may see from Eq.~\eqref{eq:transf}). For relatively small tilt angles, the principal axis of the quadrupole wake rotates toward the beam rotation, as observed in Ref.~\cite{SKWK}. The second effect is the increase in the quadrupole wake amplitude. It should be stressed that both effects cannot be easily compensated, as in the case of a pure quadrupole wake, where one may use alternating slab structures discussed in Refs.~\cite{Xiao,Lynn}. This is because both the amplitude and the skew \rd{angle} are highly dependent on the relative position of the beam with respect to the structure.


\section{Transverse Skewed Wake Potentials Generated by an Elliptic Beam} \label{section:elliptic-beam}

In this section, we consider a more realistic beam model. We adopt a model of the beam distribution suggested in Ref.~\cite{planar} and assume that the beam is centered at the origin of the slab structure. Its transverse charge distribution is given by
\begin{align}
    \rho_\perp(x_0,y_0) = \dfrac{1}{\sqrt{2\pi} \sigma_x} \exp\left(-\dfrac{x_0^2}{2\sigma_{x}^{2}}\right)\delta(y_0),
\end{align}
where \( \sigma_x \) is the beam width in the \( x \)-direction.

A beam that has a tilt angle \( \alpha \) with respect to the channel of the slab structure has the following distribution:
\begin{multline} \label{density-rotated}
     \Tilde{\rho}_\perp (x_0,y_0,\alpha)= \dfrac{1}{\sqrt{2\pi} \sigma_x} \exp\left[-\dfrac{(x_0\cos{\alpha} + y_0\sin{\alpha})^2}{2\sigma_{x}^{2}}\right] \\ \times \delta(-x_0\sin \alpha + y_0 \cos \alpha).
\end{multline}

The transverse wake potential of the tilted beam is calculated as
\begin{align} \label{eq:W}
    W(x,y,\alpha) = \iint w_{\perp}(x,y,x_0,y_0) \Tilde{\rho}_\perp (x_0,y_0,\alpha) \, dx_0 \, dy_0,
\end{align}
where \( w_\perp \) is given by Eq.~\eqref{potent}. The integral over \( y_0 \) can be easily evaluated, and Eq.~\eqref{eq:W} reduces to
\begin{align} \label{intW}
    &W(x,y,\alpha) = \dfrac{1}{\sqrt{2\pi} \sigma_x \cos{\alpha}} \nonumber\\
    &\int\limits_{-\infty}^{\infty} w_{\perp}(\psi, x_0, x_0 \tan \alpha) \exp\left[-\dfrac{x_0^2}{2\sigma_{x}^{2} (\cos{\alpha})^2}\right] \, dx_0.
\end{align}

We note that
\begin{equation} \label{1-itan}
    1 + i \tan \theta = \dfrac{\cos \theta + i \sin \theta}{\cos \theta} = \dfrac{e^{i \theta}}{\cos \theta}.
\end{equation}
Using Eq.~\eqref{potent} and Eq.~\eqref{1-itan}, Eq.~\eqref{intW} can be written as
\begin{multline} \label{W-new}
     W(x,y,\alpha) = -\dfrac{1}{\sqrt{2\pi} \sigma_x \cos{\alpha}} \\
     \int\limits_{-\infty}^{\infty} \dfrac{\tanh\left[\dfrac{\pi(x-iy)}{4a} - \dfrac{\pi x_0}{4a} \dfrac{e^{i \alpha}}{\cos \alpha}\right]}{\left\{\cosh\left[\dfrac{\pi(x-iy)}{4a} - \dfrac{\pi x_0}{4a} \dfrac{e^{i \alpha}}{\cos \alpha}\right]\right\}^2} \\
     \exp\left[-\dfrac{x_0^2}{2\sigma_{x}^{2} (\cos{\alpha})^2}\right] \, dx_0.
\end{multline}

To extract the quadrupole part from Eq.~\eqref{W-new}, we linearize \( W \) near the origin and obtain
\begin{align} \label{eq:wb-quad}
W^q(x,y,\alpha) = \frac{1}{\sqrt{2\pi} \sigma_x}\left[\frac{\pi(-x+iy)}{4a}I_1(\alpha) + I_2(\alpha) \right],
\end{align}
where
\begin{align} \label{eq:M-Int}
&I_1(\alpha) = \dfrac{1}{\cos{\alpha}} \\
&\int\limits_{-\infty}^{\infty} \dfrac{2 - \cosh\left[\dfrac{\pi x_0}{2a} \dfrac{e^{i \alpha}}{\cos \alpha}\right]}{\left\{\cosh\left[\dfrac{\pi x_0}{4a} \dfrac{e^{i \alpha}}{\cos \alpha}\right]\right\}^4} 
     \exp\left[-\dfrac{x_0^2}{2\sigma_{x}^{2} (\cos{\alpha})^2}\right] \, dx_0 \nonumber
\end{align}
and
\begin{align}
&I_2(\alpha) = \dfrac{2}{\cos{\alpha}} \\ \nonumber
&\int\limits_{-\infty}^{\infty} \dfrac{\sinh\left[\dfrac{\pi x_0}{2a} \dfrac{e^{i \alpha}}{\cos \alpha}\right]}{\left\{\cosh\left[\dfrac{\pi x_0}{4a} \dfrac{e^{i \alpha}}{\cos \alpha}\right]\right\}^{4}} 
     \exp\left[-\dfrac{x_0^2}{2\sigma_{x}^{2} (\cos{\alpha})^2}\right] \, dx_0.
\end{align}

Substituting
\begin{align}
t = \dfrac{\pi}{4a} x_0 \dfrac{e^{i \alpha}}{\cos \alpha},
\end{align}
we reduce both integrals to
\begin{align}
I_1(\alpha) = \dfrac{4a e^{-i\alpha}}{\pi} \int\limits_{\Gamma} \dfrac{2 - \cosh 2t}{(\cosh t)^4} 
     \exp\left[-\dfrac{t^2}{2 \kappa(\alpha)^2}\right] \, dt
\end{align}
and
\begin{align}
I_2(\alpha) = \dfrac{8a e^{-i\alpha}}{\pi}
\int\limits_{\Gamma} \dfrac{\sinh 2t}{(\cosh t)^4} 
     \exp\left[-\dfrac{t^2}{2 \kappa(\alpha)^2}\right] \, dt,
\end{align}
where \( \kappa \) is given by
\begin{align}
\kappa(\alpha) = \dfrac{\pi}{4} \dfrac{\sigma_x}{a} \exp(i\alpha).
\end{align}

\begin{figure}
    \centering
    \includegraphics[width=1\linewidth]{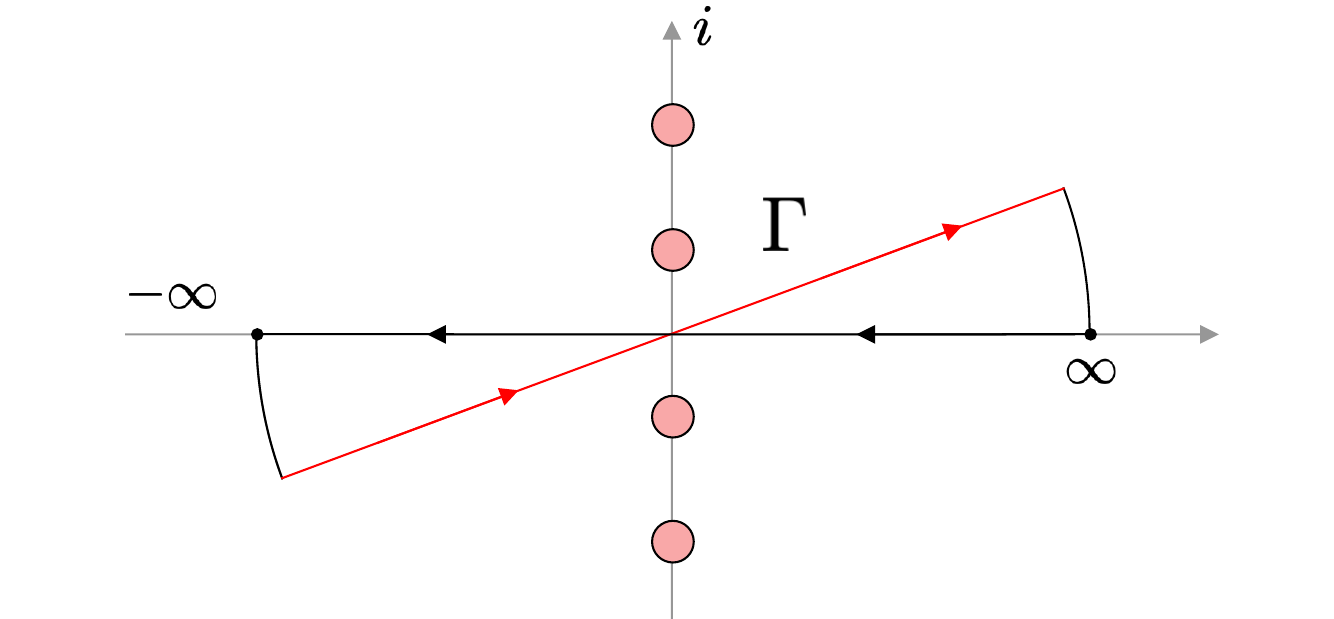}
    \caption{Integration path \( \Gamma \) (red line) and a closure of the integration path (two black arcs and a line \( x \in [-\infty, \infty] \)). Points are poles of the hyperbolic cosine - the only singularities of the functions under the integrals \( I_1 \) and \( I_2 \). Integrals over the arcs are zero; consequently, \( \int\limits_\Gamma = \int\limits_{-\infty}^{\infty} \).}
    \label{fig:6}
\end{figure}

The integration path \( \Gamma \) (see Fig.~\ref{fig:6}) is a rotated \( x \)-axis in the complex plane. We close the integration path as shown in Fig.~\ref{fig:6} and note that both functions under the integrals \( I_1 \) and \( I_2 \) are analytic on the closed path and inside the closed path. This allows us to rewrite the integrals as
\begin{align} \label{eq:I1}
I_1(\alpha) = \dfrac{4a e^{-i\alpha}}{\pi} \int\limits_{-\infty}^{\infty} \dfrac{2 - \cosh 2t}{(\cosh t)^4} 
     \exp\left[-\dfrac{t^2}{2 \kappa(\alpha)^2}\right] \, dt
\end{align}
and
\begin{align} \label{eq:I2}
I_2(\alpha) = \dfrac{8a e^{-i\alpha}}{\pi}
\int\limits_{-\infty}^{\infty} \dfrac{\sinh 2t}{(\cosh t)^4} 
     \exp\left[-\dfrac{t^2}{2 \kappa(\alpha)^2}\right] \, dt.
\end{align}

The integral \( I_2 \) is zero because the integrand in Eq.~\eqref{eq:I2} is odd. The integral \( I_1 \) can be evaluated approximately by expanding the Gaussian exponent in a series with respect to \( 1/|\kappa(\alpha)^2| \). This corresponds to the assumption of a highly elliptic beam \cite{planar}, where
\begin{align}
\varkappa \equiv \sigma_x / a \gg 1.
\end{align}
Thus, we obtain
\begin{align} \label{eq:I1-app}
I_1(\alpha) \approx \dfrac{4a}{\pi} \frac{16}{\pi^2} \left[\frac{e^{-3i\alpha}}{\varkappa^2} - \frac{2e^{-5i\alpha}}{\varkappa^4} \right].
\end{align}

From Eq.~\eqref{eq:wb-quad}, it is apparent that the skewed quadrupole wake can be expressed in terms of the regular quadrupole wake (when the tilt angle is zero, \( \alpha = 0 \)) as
\begin{align}
W^q(x,y,\alpha) = \frac{I_1(\alpha)}{I_1(0)} W^q(x,y,0).
\end{align}

Consequently, by analogy with Eq.~\eqref{eq:skang} and Eq.~\eqref{eq:skamp}, we introduce a skew angle
\begin{align} \label{eq:skanb}
\phi(\varkappa) = \dfrac{\arg[I_1(\alpha)]}{2}
\end{align}
and a scaling factor
\begin{align} \label{eq:skampb}
\lambda(\varkappa) = \dfrac{|I_1(\alpha)|}{I_1(0)}.
\end{align}

Using Eq.~\eqref{eq:I1-app}, we obtain an asymptotic expression for the scaling factor in the case of \( \varkappa \gg 1 \):
\begin{align} \label{eq:skama}
\lambda(\varkappa) \approx 1 + \frac{4 (\sin \alpha)^2}{\varkappa^2},
\end{align}
as well as for the skew angle:
\begin{align} \label{eq:skaa} 
\phi(\varkappa) \approx -\frac{3}{2} \alpha + \frac{\sin 2\alpha}{\varkappa^2}.
\end{align}

In the opposite case of \( \varkappa \ll 1 \), the integral in Eq.~\eqref{eq:M-Int} can be evaluated analytically by substituting \( \tilde{t} = x_0 / \sigma_x / \cos \alpha \):
\begin{align}
I_1(\alpha) = \sigma_x 
\int\limits_{-\infty}^{\infty} \dfrac{2 - \cosh\left[\dfrac{\pi \tilde{t}}{2} \varkappa e^{i \alpha}\right]}{\left\{\cosh\left[\dfrac{\pi \tilde{t}}{4} \varkappa e^{i \alpha}\right]\right\}^4} 
     \exp\left[-\dfrac{\tilde{t}^2}{2}\right] \, d\tilde{t}.
\end{align}
By expanding in \( \varkappa \), we obtain
\begin{align} \label{eq:I1-app2}
I_1(\alpha) \approx \sqrt{2\pi} \sigma_x \left[1 - \frac{\pi^2}{4} e^{2i\alpha} \varkappa^2 \right].
\end{align}

Using Eq.~\eqref{eq:skanb} and Eq.~\eqref{eq:I1-app2}, we find the skew angle for a beam with small ellipticity \( \varkappa \ll 1 \):
\begin{align} \label{eq:skb}
\phi(\alpha) \approx -\frac{\pi^2}{8} \varkappa^2 \sin 2\alpha.
\end{align}

For the scaling factor, from Eq.~\eqref{eq:I1-app2}, we obtain
\begin{align} \label{eq:scb} 
\lambda \approx 1 + \frac{\pi^2}{2} \varkappa^2 (\sin \alpha)^2.
\end{align}

By comparing Eq.~\eqref{eq:skb} and Eq.~\eqref{eq:scb} with Eq.~\eqref{eq:approxsk} and Eq.~\eqref{eq:approxap}, we observe that a beam with small ellipticity can be well described by the three-particle model with the following relationship between the \( d \) parameter and the beam ellipticity parameter:
\begin{align}
d = \sqrt{\dfrac{3}{2}} \varkappa a.
\end{align}

Finally, we note the following simple upper bound on the \( \varkappa \) parameter:
\begin{align} \label{eq:kpmax}
\varkappa \leq \frac{1}{3 \sin \alpha}.
\end{align}
This follows from the observation that the full half-width \( 3\sigma_x \) of the tilted beam cannot be larger than the clearance gap of \( a / \sin \alpha \) if one assumes beam transmission without particle loss.

\begin{figure}
    \centering
    \includegraphics[width=1\linewidth]{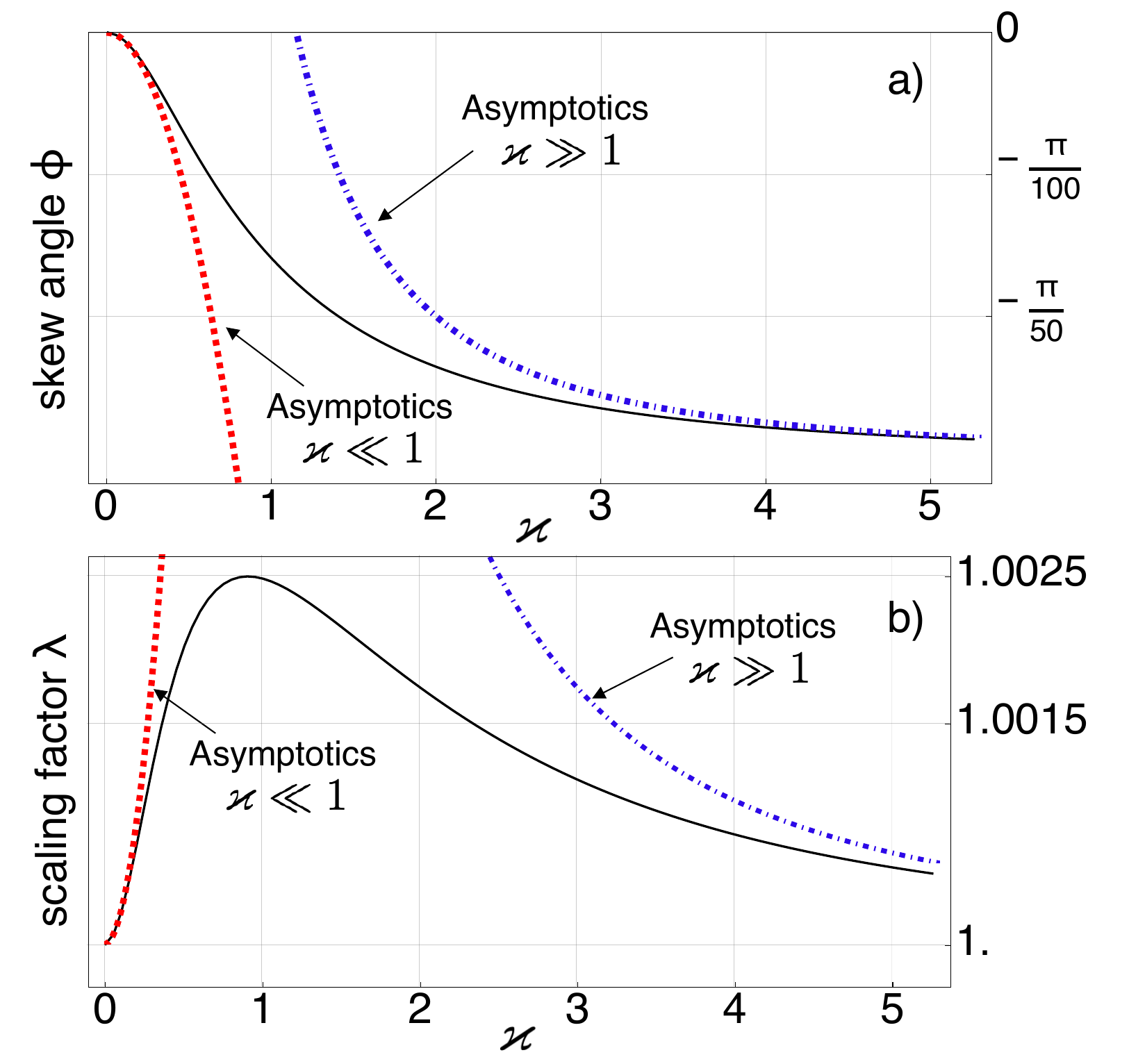}
    \caption{Skew angle \( \phi \) (panel a) and scaling factor \( \lambda \) (panel b) as functions of the beam ellipticity parameter \( \varkappa \equiv \sigma_x / a \) for a beam tilt angle \( \alpha = \pi/50 \). The maximum value for \( \varkappa_{\mathrm{max}} = 5.3 \) is estimated from the inequality in Eq.~\eqref{eq:kpmax}. Panel a): The red dashed line corresponds to Eq.~\eqref{eq:skb}, and the blue dot-dashed line corresponds to Eq.~\eqref{eq:skaa}; the black line corresponds to the exact formula in Eq.~\eqref{eq:skanb} with \( I_1(\alpha) \) given by Eq.~\eqref{eq:M-Int}. Panel b): The red dashed line corresponds to Eq.~\eqref{eq:scb}, and the blue dot-dashed line corresponds to Eq.~\eqref{eq:skama}; the black line corresponds to the exact formula in Eq.~\eqref{eq:skampb} with \( I_1(\alpha) \) given by Eq.~\eqref{eq:M-Int}.}
    \label{fig:7}
\end{figure}

In Fig.~\ref{fig:7}, we compare asymptotic formulas for the skew angle \( \phi \) and the scaling factor \( \lambda \) with the case where the master integral in Eq.~\eqref{eq:M-Int} is evaluated numerically. We observe good agreement in both limiting cases.

We would like to point out two important consequences of the above analysis.

First, the scaling factor is significant for beams with low ellipticity (\( \varkappa \leq 1 \)). As can be seen in Fig.~\ref{fig:7}, \( \lambda \) has a maximum around \( \varkappa = 1 \) and rapidly decays as \( 1/\varkappa^2 \). An additional limitation on the upper value of the scaling factor comes from the inequality in~\eqref{eq:kpmax}, which connects the maximum tilt angle and the beam width \( \sigma_x \) for a given gap \( 2a \). The wider the beam, the smaller the allowed tilt angle, and consequently the smaller the maximum value of the scaling parameter. In the example considered in Fig.~\ref{fig:7}, to ensure a clean passage of the elliptic beam with \( \varkappa = 5.3 \) without hitting the wall, one must keep the tilt angle below \( \alpha < \pi/50 \). This restriction results in a natural suppression of the amplitude effect in highly elliptic beams, provided that the beam passes through the structure without clipping.

Second, the skew effect always remains. As follows from Eq.~\eqref{eq:skaa}, the limiting value of the skew angle \( \phi \) is \rd{\( -3/2 \)} of the tilt angle \( \alpha \) for the case of an extremely elliptic beam. It must be noted that, in any realistic scenario, it is impossible to inject the beam perfectly parallel to a given surface, which means that the skew effect is always present. Due to the fact that the skew effect depends on the beam angle, it has a stochastic nature and is unlikely to be compensated for, as discussed in Refs.~\cite{Xiao,Lynn} for the pure quadrupole wake. Consequently, one may expect an instability connected with the skew wake effect.

\section{Conclusion}

In this work, we have investigated the skew wake effect in slab structures, focusing on its dependence on beam geometry and tilt angle. By employing both a three-particle model and a more realistic elliptic beam model, we derived analytical expressions for the skew angle \( \phi \) and scaling factor \( \lambda \) of the transverse wake potential. Our analysis reveals that these quantities are strongly influenced by the beam's tilt angle \( \alpha \) and ellipticity parameter \( \varkappa = \sigma_x / a \). For highly elliptic beams \bl{with the Gaussian transverse distribution} (\( \varkappa \gg 1 \)), the skew angle approaches \( \phi \approx -\frac{3}{2} \alpha \), while the scaling factor \( \lambda \) is suppressed as \( \lambda \approx 1 + \frac{4 (\sin \alpha)^2}{\varkappa^2} \). In contrast, for beams with low ellipticity (\( \varkappa \ll 1 \)), the skew angle scales as \( \phi \approx -\frac{\pi^2}{8} \varkappa^2 \sin 2\alpha \), and the scaling factor grows as \( \lambda \approx 1 + \frac{\pi^2}{2} \varkappa^2 (\sin \alpha)^2 \).

The sensitivity of the wake amplitude to the skew effect is most pronounced for beams with low ellipticity (\( \varkappa \leq 1 \)), where the scaling factor \( \lambda \) reaches its maximum. However, for highly elliptic beams (\( \varkappa \gg 1 \)), the scaling factor is naturally suppressed due to the constraint \( \varkappa \leq \frac{1}{3 \sin \alpha} \), which ensures the beam does not clip. Despite this suppression, the skew effect, characterized by the rotation of the quadrupole wake's principal axis, persists even for highly elliptic beams. The skew angle \( \phi \) is \rd{not} always proportional to the tilt angle \( \alpha \). \rd{We stress that the proportionality \( \phi \approx -\frac{3}{2} \alpha \)  holds only in the limit of highly elliptical beam \( \varkappa \gg 1 \) when the term depending on the ellipticity $\varkappa$ can be fully neglected.}

The basic theory we used is independent of the specific nature of the wake field (as demonstrated in Refs.\cite{MyPRL,mySTAB,Brendan:2020}), leading to the conclusion that the skew wake is a purely geometric phenomenon. It manifests itself in any interaction with asymmetric beams and results from imperfect beam alignment, which is inherently random. Consequently, the skew effect is likely to cause instabilities or emittance growth, especially in applications that require precise control of beam dynamics, such as colliders and wakefield accelerators.

The stochastic nature of the skew wake effect poses significant challenges for beam control and compensation. Unlike the case of a pure quadrupole wake, where alternating slab structures can mitigate the effect, the skew wake's dependence on random beam misalignment makes it difficult to suppress. This underscores the importance of advanced beam diagnostics and feedback systems to manage the impact of the skew wake effect in practical applications.

Looking ahead, further experimental studies are needed to investigate this effect, particularly for beams with extreme ellipticity and small tilt angles. These efforts will be essential for the design and operation of future accelerators and colliders, where understanding and controlling the skew wake effect will play a crucial role in achieving optimal beam performance.

In summary, our results show that the skew wake effect is an intrinsic feature of transversely asymmetric beams in slab structures. While the scaling factor \( \lambda \) is naturally suppressed for highly elliptical beams, the skew angle \( \phi \) remains significant and can lead to beam instabilities. These results highlight the importance of considering the skew wake effect in the design and operation of future accelerators and colliders, and emphasize the need for continued research and innovation in this area.

\begin{acknowledgments}
The work was supported by the Foundation for the Advancement of Theoretical Physics and Mathematics ``BASIS" $\#$22-1-2-47-17. 
\end{acknowledgments}

\appendix
\section{Connection of the rotation of the coordinate system to the rotation of the quadrupole vector field}\label{app:separ}
We start from the rotated quadrupole field in a form
\begin{align}
\label{eq:Frot}
\vec F_\perp = \begin{pmatrix}
 \cos \alpha & \sin\alpha \\
  -\sin \alpha& \cos\alpha
\end{pmatrix}\begin{pmatrix}
    -x \\ y
\end{pmatrix}.
\end{align}
The corresponding potential for this field reads
 \begin{equation} \label{potentialpov}
    U(x,y) =  \dfrac{\left(x^2-y^2\right) \cos (\alpha ) }{2}- xy\sin (\alpha ).
\end{equation}
Now we consider a potential for the regular (non rotated) quadrupole vector field 
\begin{align}
\label{eq:nonrotpot}
U_0(x',y')=\frac{x'^2-y'^2}{2},
\end{align}
and consider a \rd{counter} clockwise rotation of the coordinate system
\begin{align}
\label{eq:crot}
\begin{pmatrix}
    x \\ y
\end{pmatrix} = \begin{pmatrix}
 \cos \phi & -\sin\phi \\
  \sin \phi& \cos\phi
\end{pmatrix}\begin{pmatrix}
    x' \\ y'
\end{pmatrix}.
\end{align}
By plugin Eq.\eqref{eq:crot} into Eq.\eqref{eq:nonrotpot} we get
\begin{align}
\label{eq:rotpot}
  U_0(x,y)=\frac{x^2-y^2}{2}\cos 2\phi+xy\sin2\phi  
\end{align}
By comparing Eq.\eqref{potentialpov} and Eq.\eqref{eq:rotpot} we conclude 
\begin{align}
\phi=-\frac{1}{2}\alpha.
\end{align}
This means that the rotation angle in the vector field rotation matrix is minus twice the coordinate system rotation angle.

\bibliographystyle{ieeetr}
\bibliography{refs}

\end{document}